\documentclass[lettersize,journal]{IEEEtran}

\usepackage{amsmath}
\usepackage{amsfonts}
\usepackage{algorithmic}
\usepackage{array}
\usepackage{booktabs}
\usepackage{graphicx}
\usepackage{indentfirst}
\usepackage{url}
\usepackage{siunitx}
\usepackage{stfloats}
\usepackage{enumitem}
\usepackage{subfigure}
\usepackage{xcolor}
\usepackage{textcomp}
\usepackage{verbatim}
\usepackage{balance}

\newcommand{\tabref}[1]{{Table \ref{#1}}}
\newcommand{\secref}[1]{{Section \ref{#1}}}

\newcommand{\figref}[1]{{Figure \ref{#1}}}

\DeclareMathOperator{\sign}{sign}

\begin{document}
\title{Pitch Contour Exploration Across Audio Domains: A Vision-Based Transfer Learning
Approach}
\author{Jakob~Abe{\ss}er,~\IEEEmembership{Member,~IEEE,} Simon~Schw{\"a}r, Meinard~M{\"u}ller,~\IEEEmembership{Fellow,~IEEE}
\thanks{Jakob~Abe{\ss}er, Simon~Schw{\"a}r, and Meinard~M{\"u}ller are with International Audio Laboratories Erlangen, Germany.}
\thanks{Jakob~Abe{\ss}er is further with Fraunhofer IDMT, Ilmenau, Germany.}
}

\markboth{Journal of \LaTeX\ Class Files,~Vol.~18, No.~9, September~2020}%
{How to Use the IEEEtran \LaTeX \ Templates}

\maketitle

\begin{abstract}
This study examines pitch contours as a unifying semantic construct prevalent across various audio domains including music, speech, bioacoustics, and everyday sounds. Analyzing pitch contours offers insights into the universal role of pitch in the perceptual processing of audio signals and contributes to a deeper understanding of auditory mechanisms in both humans and animals. Conventional pitch-tracking methods, while optimized for music and speech, face challenges in handling much broader frequency ranges and more rapid pitch variations found in other audio domains. This study introduces a vision-based approach to pitch contour analysis that eliminates the need for explicit pitch-tracking. The approach uses a convolutional neural network, pre-trained for object detection in natural images and fine-tuned with a dataset of synthetically generated pitch contours, to extract key contour parameters from the time--frequency representation of short audio segments. A diverse set of eight downstream tasks from four audio domains were selected to provide a challenging evaluation scenario for cross-domain pitch contour analysis.
The results show that the proposed method consistently surpasses traditional techniques based on pitch-tracking on a wide range of tasks.
This suggests that the vision-based approach  establishes a foundation for 
comparative studies of pitch contour characteristics across diverse audio domains.
\end{abstract}

\begin{IEEEkeywords}

\end{IEEEkeywords}

\section{\label{sec:intro} Introduction}

\begin{figure*}[t]
\includegraphics[width=\textwidth]{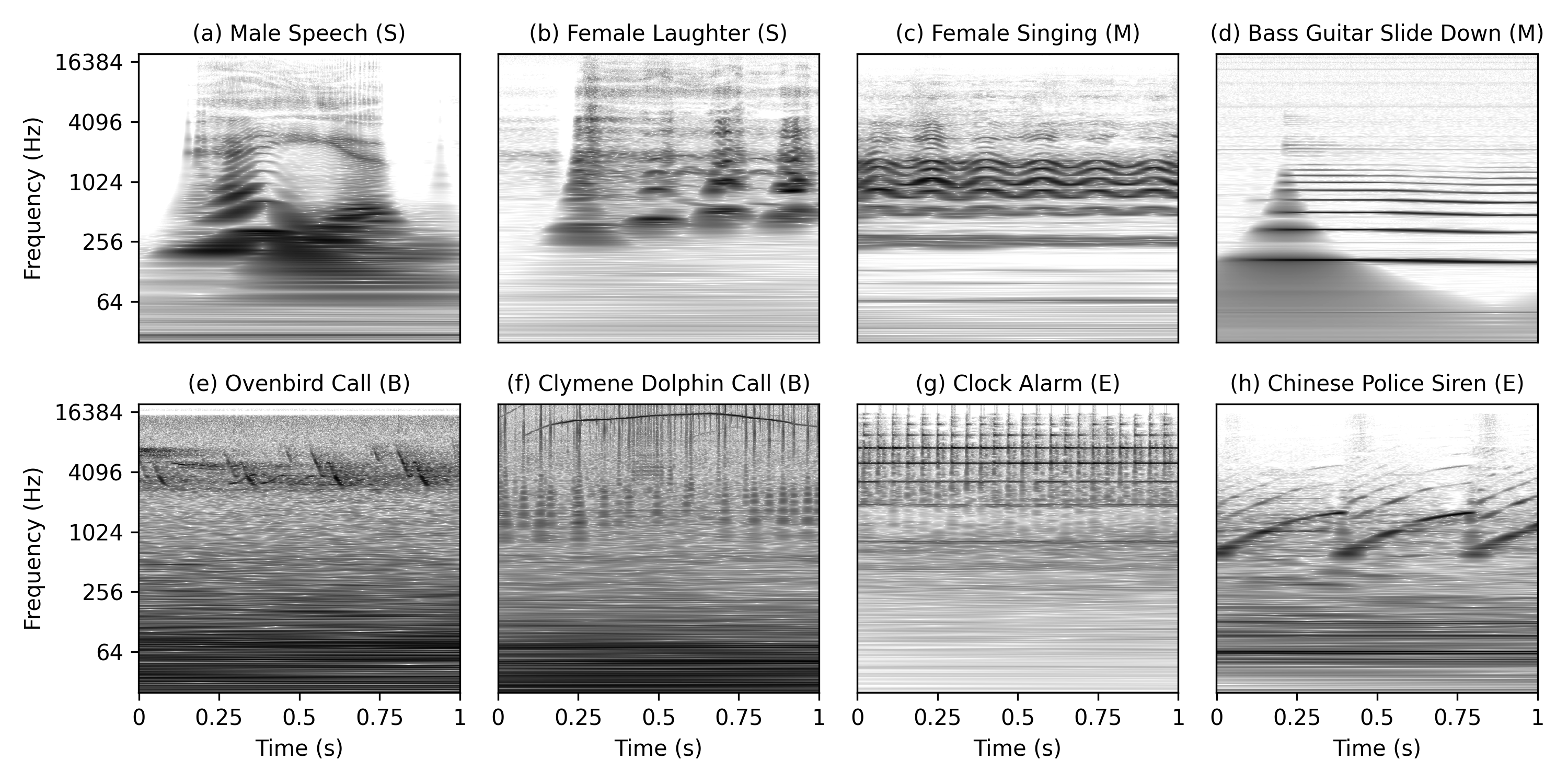}
\caption{\label{fig:downstream_task_examples}
{Excerpts of pitch contours selected from four different audio domains covering speech (S), music (M), bioacoustics (B), and everyday sounds (E). Each pitch contour is displayed as a one-second Constant-Q spectrogram with log-scaled magnitude.
}
}
\end{figure*}

Pitch is an important perceptual attribute in human and animal auditory processing and describes the subjective sensation of frequency. 
Pitch is most commonly associated with the lowest frequency that significantly contributes to the perception of a sound, known as its fundamental frequency (F0).
In natural sounds, pitch rarely remains constant, but rather varies over time.
This is illustrated in \figref{fig:downstream_task_examples} for several short examples from the audio domains speech (S), music (M), bioacoustics (B), and everyday sounds (E).
Each sound is represented as a Constant-Q spectrogram, highlighting the F0 and its overtones as prominent horizontal, diagonal, or arc-shaped contours.
Examples include low-pitched sounds such as male speech (a) or a bass guitar note (d), as well as high-pitched sounds such as bird calls (e), a dolphin call (f), and an alarm clock (g). 
In many cases, information is encoded in the variations of pitch over time on multiple time scales. For example, in music performances, periodic short-term pitch variations such as vibrato (c) are used as an expressive element, while a long-term sequence of distinct pitches may convey a melody.
Alarm sounds, such as clock alarms (g) and sirens (h), are designed with repeated cycles of alternating, increasing, or decreasing pitch to create a distinctive auditory signal that is easily and quickly identifiable even in noisy environments. 
To study pitch variations over time, we define the temporal sequence of F0 values as the pitch contour (PC). This concept is then used to compare sounds across different audio domains.


Before PCs can be analyzed, they are typically extracted with a pitch-tracking algorithm by calculating frame-wise estimates of the local pitch. 
Initially, pitch-tracking methods focused on signal processing techniques for speech speech (e.\,g., \cite{Rabiner_1977_ACFPitchDetection_TASLP}) and music (e.\,g., \cite{CamachoH08_SawtoothWaveform_JASA}) analysis.
More recently, deep neural networks, such as convolutional neural networks (CNNs), are used to estimate PCs directly from 
a time- or frequency-domain representation of the audio signal 
in an end-to-end manner (e.\,g., \cite{Wook_2018_CREPE_ICASSP, Gfeller_2020_SPICE_TASLPS}).
The extracted PCs can then be further analyzed for different tasks, such as genre or singing style classification using hand-crafted features \cite{Bittner_2015_MelodyContours_ISMIR, Salamon_2012_MelodyFeatures_ICASSP} or learned feature representations \cite{Abesser:2019:ICASSP, Han_2023_FindingTori_ISMIR}.

Despite recent advancements, most pitch-tracking algorithms are tailored to specific domains, relying on explicit assumptions to improve the robustness of pitch estimation.
For example, some pitch trackers limit the maximum rate of change of PCs to identify the most likely pitch estimate for successive frames in a post-processing step \cite{Mauch_2014_pYin_ICASSP}, which can be suitable for music signals, but inappropriate for, e.g., bird songs \cite{Stowell_2014_FrequencyModulationBird_MEE}.
Data-driven approaches, in contrast, typically eliminate the need for explicit model assumptions but rely heavily on extensive domain-specific training data to achieve reliable results for PC analysis.

In this article, we advance unified PC analysis and foster a deeper understanding of PCs across diverse audio domains.
As a first contribution, we introduce the Synthetic Pitch Contours (SPC) dataset, which encompasses seven PC types with distinct characteristics, including vibrato, glissando, and pitch bends (see \secref{sec:spc}).
The dataset is designed to encompass diverse use cases through a wide parameter range and is motivated by finding common PC descriptors for different audio domains.
As a second contribution, we propose a novel method for PC analysis that eliminates the need for an explicit pitch-tracking step (\secref{sec:e2e_pc_modeling} to \secref{sec:evaluation}). Our approach leverages a deep neural network pre-trained for object detection in natural images \cite{Sandler_2018_MobileNetV2_CVPR}, which we fine-tune (using the SPC dataset) to estimate contour parameters and the overall shape of PCs from the time--frequency representation of short audio clips.
Finally, as a third contribution, we evaluate the effectiveness of this approach across multiple downstream classification tasks related to speech, music, bioacoustics, and everyday sounds (\secref{sec:downstream_tasks}).

\begin{figure*}[t]
\includegraphics[width=1\textwidth]{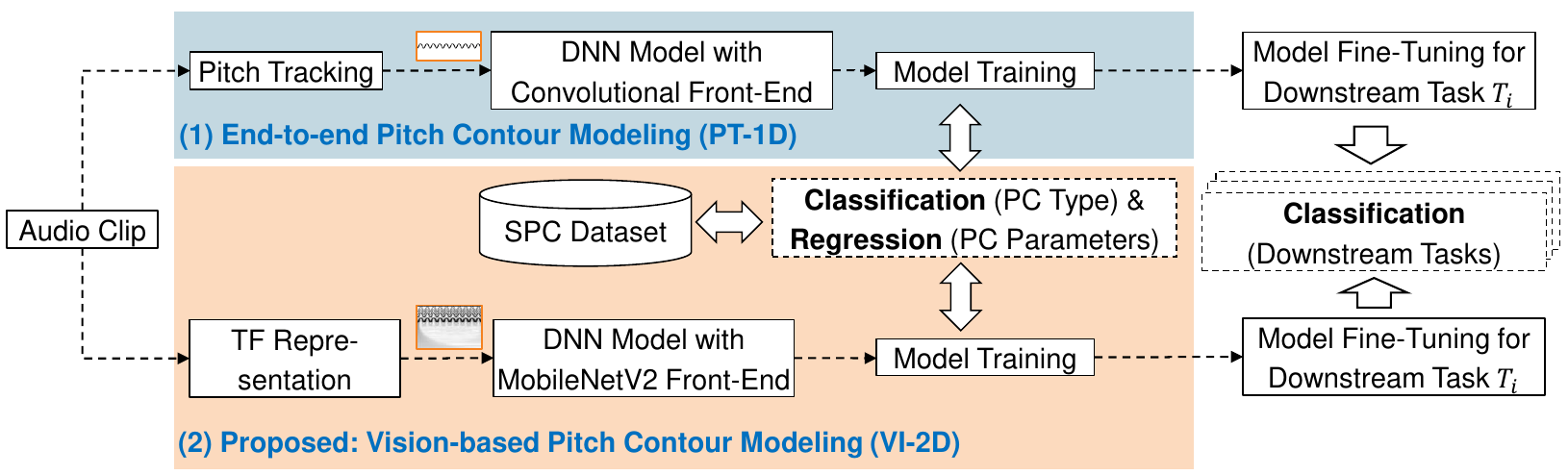}
\caption{\label{fig:overall_flowchart}
{Two approaches for PC analysis. (1) End-to-end approach (\texttt{PT-1D}) with PCs extracted using a pitch-tracking algorithm and being processed by a deep neural network (DNN) model with a trainable convolutional front-end. (2) Proposed vision-based approach  (\texttt{VI-2D}) with PCs being captured as time--frequency (TF) representations  of audio clips and processed by a DNN model using a MobileNetV2 front-end, which has been pre-trained on ImageNet. In both approaches, pre-trained models are later fine-tuned for different downstream classification tasks.}}

\end{figure*}

\figref{fig:overall_flowchart} provides a methodological overview: A baseline approach (PT-1D) is illustrated in the blue box (top). Here, a pitch-tracking algorithm that relies on traditional signal processing (see \secref{sec:pitch_tracking} for details) is used to explicitly extract the PC from an audio clip, represented as a fundamental frequency (F0) trajectory.
The PC is then processed by a deep neural network (DNN) with a convolutional front-end (see \secref{sec:nn_architecture_e2e}).
In the second (proposed) approach (\texttt{VI-2D}) shown in the orange box (bottom), an audio clip is first converted into a two-dimensional time--frequency representation (see \secref{sec:tf_representations}). This representation is then processed by a DNN with a \textit{MobileNetV2} \cite{Sandler_2018_MobileNetV2_CVPR} front-end (see \secref{sec:nn_architecture_vb}).

Before diving into these contributions, \secref{sec:pcs_across_domains} gives an overview of related research on pitch contours across various domains. Conclusions and perspectives are presented in \secref{sec:conclusion}.

\section{\label{sec:pcs_across_domains} Pitch Contours Across Domains}

In speech communication, PCs convey important information about intonation, prosody, and expressiveness \cite{Scherer_2003_VocalExpression_HAS}.
PCs extracted from speech signals include sequences of stable pitch segments, which are suitable for intonation analysis, and intra-syllabic pitch variations such as increasing and decreasing glissandi \cite{Allesandro_1995_PitchContourStylization_CSL}.
An important area of study is to develop models that connect PCs to the speaker's emotions \cite{Rabiei_2014_SpeechEmotion_JBR}, utilizing fundamental pitch metrics like mean and range as examples \cite{Busso_2009_SpeechEmotion_TASLP}.

In music, pitch is essential as it establishes the tonal basis of a composition and facilitates the creation of melodies and harmonies. 
Vocalists deliberately use pitch modulations as an effective means of musical expression in various music styles ranging from Western art songs \cite{Prame_1997_VibratoLyricSinging_JASA}, over classical Indian Hindustani and Carnatic music \cite{Vidwans_2012_IndianClassicalVocal_CompMusic, Gupta_2011_OrnamentationsIndianSinging_BOOK}, to other non-Western singing traditions \cite{Proutskova_2023_VocalNotes_ISMIR}.
For example, singers convey emotions through periodic frequency modulations (vibrato) or continuously rising frequency in the direction of a target tone (glissando). \cite{Sundberg_1998_SingingExpressivity_LPV}.
In the performance of string instruments such as violin \cite{Barbancho_2009_ViolinTranscription_ICASSP}, electric guitar \cite{Kehling:2014:DAFX}, bass guitar \cite{Abesser_2011_BassModulation_AES}, or the Turkish ney \cite{Oezaslan_2012_Embellishments_ISMIR}, expression styles such as pitch bends, slides (glissandi) and vibrato are widely used.
The exact execution of these techniques often depends on the musical context and the personal style of the performer or singer \cite{Abesser_2017_ScoreInformedJazzAnalysis_TASLP}.

As an interesting analogy, the contour concept plays an important role not only for PCs on a short time scale within individual musical notes, but also for melodic contours as sequences of multiple notes, where the succession of pitches can form characteristic contours on a larger time scale, such as ascending or descending melodies and melodic arches \cite{Adams_1976_MelodicContourTypology_ETH, Huron_1996_MelodicArch_CIM, Frieler:2014:FMA, Pfleiderer:2017:BOOK}.

Human pitch perception differs significantly from that of animals. 
While humans can typically perceive sounds in the range of 20~Hz and 20~kHz , they are particularly sensitive to frequencies between 2~kHz and 5~kHz, a range essential for speech comprehension.
Humans can detect and differentiate pitch changes and contours even in complex acoustic environments and inharmonic sounds \cite{McPherson_2017_PitchPerceptionDiversity_NHB}.
Animals, on the other hand, exhibit diverse hearing ranges that can be significantly broader or narrower than those of humans, extending into infrasound or ultrasound frequencies. Unlike humans, animals utilize sound not only for communication but also for critical survival tasks such as navigation and prey localization, as exemplified by bats' use of ultrasound for echolocation.

Birds, in particular, use calls and songs for species identification, mating, and territorial defense \cite{Kroodsma_2005_BirdSong_BOOK}.
Their vocalizations are composed of elements such as pitch glides, sudden jumps, and rapid trills \cite{Beckers_2003_BirdFreqAmplModulation_JEB}.
The modulation frequency range in bird calls and animal vocalizations in general varies widely between species.
Cetaceans, such as whales and dolphins, also exhibit species-specific PCs.
For example, whale songs contain stable tonal components with slow frequency sweeps \cite{McDonald_2006_WhaleSong_JCRM}.
In this context, automatic species classification methods leveraging CNNs have been developed to recognize characteristic pitch contour (PC) patterns \cite{Allen_2021_WhaleSongCNN_FMS, Kather_2024_WhaleSong_JASA, Bouffaut_2019_BlueWhale_PHD}. 

Finally, repetitive notification sounds, such as alarms or sirens from emergency vehicles, feature distinct PCs \cite{Marchegiani_2022_Sirens_TITS}. These sounds are designed to remain audible even in complex and noisy environments, such as urban soundscapes \cite{Cantarini_2021_Sirens_ISPA}.

\section{\label{sec:spc} Synthetic Pitch Contour Dataset}

This section introduces the Synthetic Pitch Contour (SPC) dataset\footnote{The SPC dataset will be made openly available on the Zenodo platform alongside this article.}, comprising \num{3500}  one-second audio clips with synthetically generated PCs.
The dataset includes seven types of PCs, which are inspired by the most common PC shapes that can be observed in different audio domains.
The dataset includes 500 audio clips for each PC type.
While ``real-life'' PCs are naturally much more complex, the one-second audio clips of the SPC datasets provide a meaningful approximation of the local shape within longer PCs. The SPC dataset spans a wide range of parameters representing PCs from various audio domains such as speech, music, bioacoustics, and environmental sounds. 
The SPC dataset consists of audio clips with individual fundamental frequency (F0) contour annotations and PC parameters and will be made publicly available alongside this article.

\subsection{\label{sec:spc_pitch_contour_types} Pitch Contour Types}

\begin{figure}[t]
\begin{center}
\includegraphics[width=.4\textwidth]{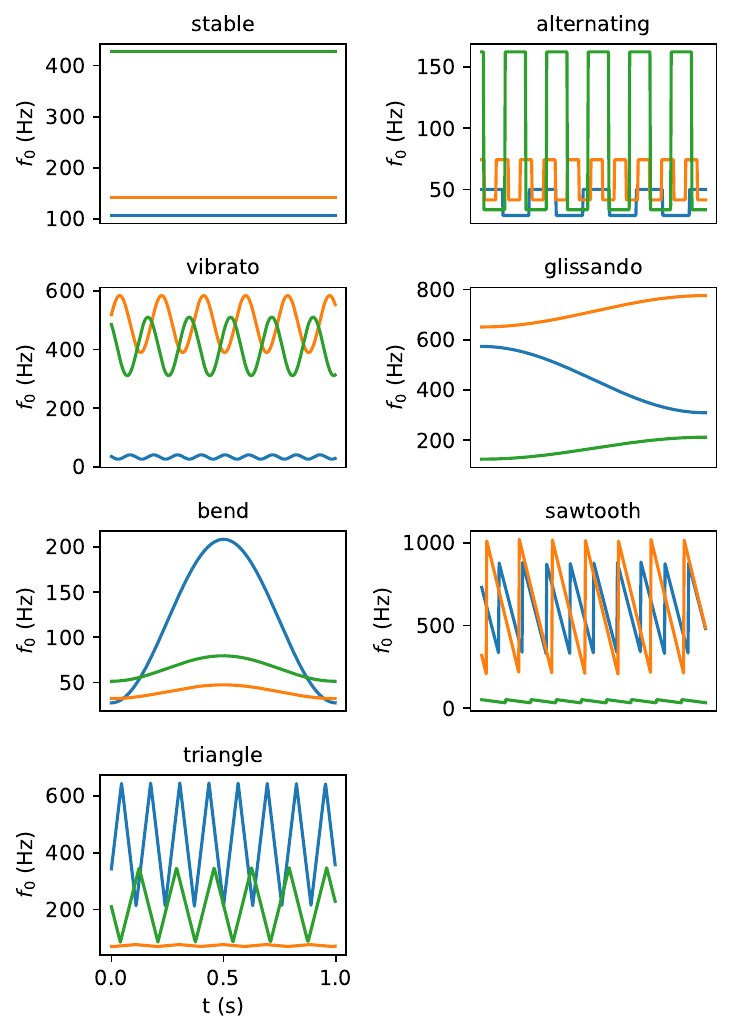}
\caption{\label{fig:contour_examples}Three example contours for each of the seven PC types.
}
\vspace{-1em}
\end{center}
\end{figure}

Following a holistic approach, we define seven types of PCs based on their characteristic F0 contour. We consider these types to represent prototypical PC shapes, which can be observed across different audio domains. \figref{fig:contour_examples} illustrates three examples for each contour type.
\begin{itemize}[itemsep=0pt,topsep=4pt]
    \item[(1)] \textit{Stable:} PCs maintain a constant F0.
    \item[(2)] \textit{Alternating:} PCs alternate periodically between two fixed F0 values, as seen in some emergency vehicle sirens or bird calls.
    \item[(3)] \textit{Vibrato:} PCs exhibit sinusoidal fluctuations in F0.
    \item[(4)] \textit{Glissando:} PCs feature a continuous transition between an initial and a target F0, either increasing or decreasing.
    \item[(5)] \textit{Bend:} PCs combine a rising and falling contour, returning to the initial F0.
    \item[(6)] \textit{Sawtooth:} PCs show a periodic, ramp-like increase in frequency, followed by an abrupt drop to the initial F0.
    \item[(7)] \textit{Triangle:} PCs resemble sawtooth contours but alternate between steady increases and decreases in frequency.
\end{itemize}

\subsection{Pitch Contour Model}
\label{sec:contour_model}

We represent each PC as a sequence of fundamental frequency (F0) values ${f_0 \in \mathbb{R}^{L}}$ over $L \in \mathbb{N}$ time frames indexed by ${n \in \left[1:L\right]:=\left\{1, 2, \ldots, L \right\}}$.
Following a unified approach for all PC types, we model $f_0$
as a base frequency $f_\mathrm{b} \in \mathbb{R}$ (in Hz) modulated by a periodic function 
$\psi: \mathbb{R}^+ \mapsto \left[-1, 1\right]$ with $\psi(x) \in \left[-1,1\right]$ and $\psi(x) \equiv \psi(x+1)$ for all $x \in \mathbb{R}^+$.
The modulation 
\begin{equation}
    f_0(n) := f_\mathrm{b} \cdot \left( 1 + 2^{\Delta_f / 1200} \cdot \psi\left(  f_\mathrm{m} \cdot T \cdot \frac{n}{L} + \phi \right)\right) 
\end{equation}
can be controlled by the modulation extent $\Delta_f \in \mathbb{R}_+$ (in cents), 
the modulation frequency $f_\mathrm{m} \in \mathbb{R}_+$ (in Hz), the overall PC duration $T \in \mathbb{R}_+$ (in seconds),
 and the phase $\phi \in \left[0,1\right]$.

Given the seven PC types introduced in \secref{sec:spc_pitch_contour_types}, we use the sine function 
\begin{equation}
\psi_\mathrm{sin}(x) := \sin(2\pi x),    
\end{equation}
to generate \textit{vibrato}, \textit{bend}, and \textit{glissando} PCs, 
the square-wave function
\begin{equation}
\psi_\mathrm{sqr}(x) := \sign \left( \sin(2\pi x)\right),
\end{equation}
to generate \textit{stable} and \textit{alternating} PCs,
the triangle wave function
\begin{equation}
\psi_\mathrm{tri}(x) = 2\left| 2 \left(x - \bigg\lfloor x + \frac{1}{2}\bigg\rfloor \right)\right|-1,
\end{equation}
to generate \textit{triangle} PCs,
and the sawtooth wave function
\begin{equation}
\psi_\mathrm{saw}(x) = 2 \left(x - \bigg\lfloor x + \frac{1}{2}\bigg\rfloor \right).
\end{equation}
to generate \textit{sawtooth} PCs.
Finally, we can time-reverse $f_0$ to generate both ascending and descending PCs for the non-symmetric contour types \textit{glissando} and \textit{sawtooth}.

\begin{table}[t]
\caption{\label{tab:constraints}Choice of periodic function $\psi$ and sampling ranges of modulation extent $\Delta_f$, modulation frequency $f_\mathrm{m}$, and phase $\phi$ for different PC types.}

\begin{tabular}{lccccc}
\toprule
\textbf{PC Type} & $\psi$ &  $\Delta_f$ & $f_\mathrm{m}~\left[\mathrm{Hz}\right]$ & $\phi$ \\
\midrule
Stable & $\psi_\mathrm{sqr}$ & $0$ & $1$ & $0$\\
Alternating & $\psi_\mathrm{sqr}$ & ${\sim \mathcal{U}(0, 1200)}$ & $\sim \mathcal{U}_\mathrm{log}(1, 50)$ & $\sim \mathcal{U}(0,1)$\\
Vibrato & $\psi_\mathrm{sin}$ & ${\sim \mathcal{U}(0, 1200)}$ & $\sim \mathcal{U}_\mathrm{log}(5, 100)$ & $\sim \mathcal{U}(0,1)$\\
Glissando & $\psi_\mathrm{sin}$ & ${\sim \mathcal{U}(0, 1200)}$ & $0.5$ & $-0.25$\\
Bend & $\psi_\mathrm{sin}$ & ${\sim \mathcal{U}(0, 1200)}$ & $1$ & $-0.25$ \\
Sawtooth & $\psi_\mathrm{saw}$ & ${\sim \mathcal{U}(0, 1200)}$ & $ \sim \mathcal{U}_\mathrm{log}(5, 100)$ & $\sim \mathcal{U}(0,1)$\\
Triangle & $\psi_\mathrm{tri}$ & ${\sim \mathcal{U}(0, 1200)}$ & $ \sim \mathcal{U}_\mathrm{log}(5, 100)$ & $\sim \mathcal{U}(0,1)$\\
\bottomrule
\end{tabular}
\end{table}

\subsection{\label{sec:parameter_sampling} 
Parameter Space for  Pitch Contour Generation}

This section outlines a parameter search space for synthesizing different PCs in the SPC database.
Each PC in the SPC dataset has a duration of $T=\SI{1}{s}$.
For example, a short two-note fragment of a vocal melody might be interpreted as a sequence of shorter PCs, such as a \textit{stable} PC, followed by a \textit{glissando}, and then another \textit{stable} PC. We consider a PC duration of 1 second sufficiently short to reasonably assume a consistent PC type.

In the following, we describe how the PC parameters are randomly sampled to generate the SPC dataset. As summarized in \tabref{tab:constraints}, different constraints are applied during sampling depending on the PC type. Two sampling approaches are used:

\begin{itemize}
\item[1.] Linear Sampling: $v \sim \mathcal{U}(v_\mathrm{min}, v_\mathrm{max})$, where the variable $v$ is drawn from a uniform distribution within in the range $\left[v_\mathrm{min}, v_\mathrm{max} \right]$.
\item[2.] Logarithmic Sampling: $v \sim \mathcal{U}_\mathrm{log}(v_\mathrm{min}, v_\mathrm{max})$, which involves two steps. First, sampling from a uniform distribution as ${\Tilde{v} \sim \mathcal{U}(\log v_\mathrm{min}, \log v_\mathrm{max})}$. Second, re-mapping the sampled value as $v := \exp \Tilde{v}$.
\end{itemize}

\begin{figure}[t]
\includegraphics[width=.5\textwidth]{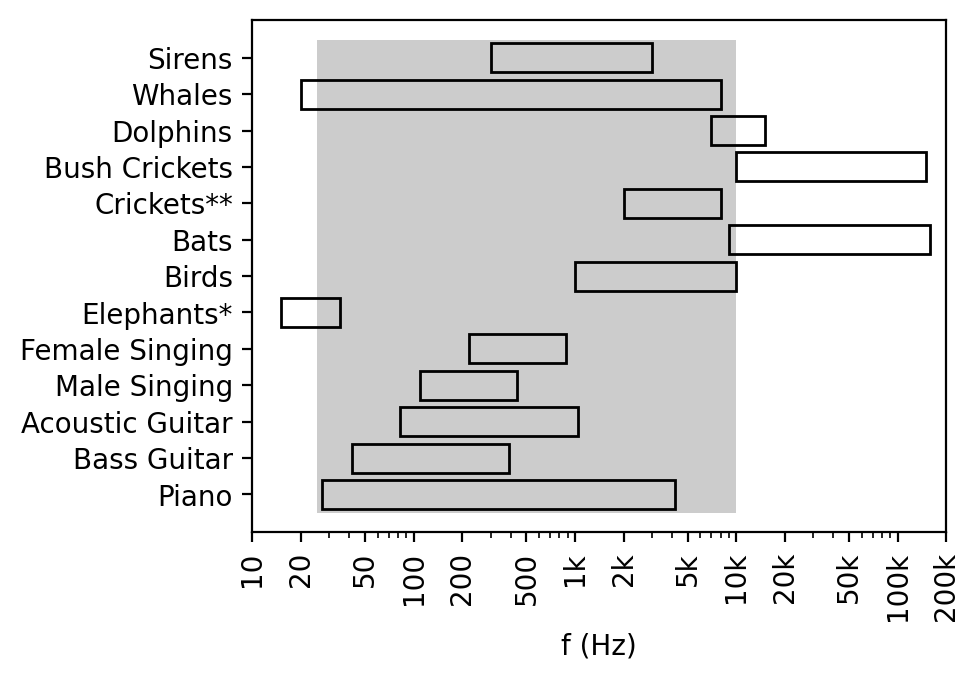}
\caption{\label{fig:frequency_ranges}{Common fundamental frequency ranges across everyday sounds, animal vocalizations, and musical instruments.
The frequency range between \SI{25}{Hz} and \SI{10}{kHz}, which is used to sample the base frequency values $f_\mathrm{b}$ in the SPC dataset, is marked as gray rectangle.} ($^*$infrasonic call, $^{**}$low frequency calls) }
\end{figure}

\textbf{Base Frequency \& Modulation Extent:} \figref{fig:frequency_ranges} illustrates typical fundamental frequency ranges of sounds from the audio domains of music, bioacoustics, and everyday sounds.
In the SPC dataset, we consider a frequency range 
between \SI{25}{Hz} and \SI{10}{kHz} to cover sounds from different audio domains with characteristic PCs, such as sirens, marine mammals (whales, dolphins), birds, singing, as well as different musical instruments.
When determining a meaningful sampling range for the modulation extent $\Delta_f$, we take into account that most intervals in both birdsongs \cite{Marler_2004_BirdSong_BOOK} and Western melodies \cite{Lerdahl_1983_TonalMusic_BOOK} typically do not exceed one octave. 
For \textit{stable} PCs, which do not exhibit frequency modulation, $\Delta_f$ is set to zero.
For all other PC types, we repeatedly sample the base frequency as ${f_\mathrm{b} \sim \mathcal{U}_\mathrm{log}(25, 10000)}$ and the modulation range as ${\Delta_f \sim \mathcal{U}(0, 1200)}$ until the entire PC lies within the targeted frequency range and satisfies the constraints:
\begin{equation}
     f_\mathrm{b} - \Delta_f \geq 25 \text{ and }  f_\mathrm{b} + \Delta_f \leq 10000.
\end{equation}

\textbf{Modulation Frequency \& Phase:} For the periodic \textit{vibrato}, \textit{sawtooth}, and \textit{triangle} PCs, we sample the modulation frequency $f_\mathrm{m}$ to range from slow frequency modulations in singing voice around \SI{5}{Hz} \cite{Titze_1994_VoiceProduction_BOOK} up to fast frequency modulations in bird calls of up to \SI{100}{Hz} \cite{Kroodsma_2005_BirdSong_BOOK}.
We use a logarithmic sampling approach as $f_\mathrm{m} \sim \mathcal{U}_\mathrm{log}(5,100)$ to emphasize slower modulations, which are more common in speech and music.
For the \textit{alternating} PCs, we take into account the highest note repetition frequencies reported in music of around \SI{25}{Hz} \cite{Collins_2002_Virtuosity_MAXIS} as well as in calls of the bunting species of up to \SI{48}{Hz}   \cite{Podos_1997_SongBirdTrills_EV}, and sample the modulation frequency as $f_\mathrm{m} \sim \mathcal{U}_\mathrm{log}(1, 50)$.
For the PC  \textit{type bend}, we use one full wave of a negative cosine function ($f_\mathrm{m}=\SI{1}{Hz}$, 
$\phi=-\frac{1}{4}$), while for the \textit{glissando}, we use an increasing half wave of the negative cosine function ($f_\mathrm{m}=\SI{0.5}{Hz}$, $\phi=-\frac{1}{4}$), which can also be time-reversed  as discussed in \secref{sec:contour_model} to implement a decreasing glissando.
In general, we randomly sample the phase as ${\phi \sim \mathcal{U}(0, 1)}$ for all PC types except for 
\textit{glissando} and \textit{bend} PCs ($\phi=-\frac{1}{4}$) and \textit{stable} PCs ($\phi=0$). 

\subsection{\label{sec:audio_synthesis} Audio Synthesis}

Given a PC $f_0 \in \mathbb{R}^L$, we synthesize an audio signal ${x \in \mathbb{R}^M}$ of length $M$ samples (indexed by $m$) using the additive sinusoidal model, where $K$ partials (indexed by $k \in \left[ 1:K\right]$) are superimposed as
\begin{equation}
    x(m) = \sum_{k=1}^{K} a_k \cdot \sin \bigg( \frac{2 \pi k}{f_\mathrm{s}} \sum_{n = 0}^m  f_0(\bar{n}) \bigg).
\end{equation}
We use a sampling rate of $f_\mathrm{s} = \SI{48}{kHz}$, which results in $M = 48000$ for a one-second audio clip.
The corresponding frame index of the PC is computed as ${\bar{n} =  }$.
We define the partial amplitudes as
\begin{equation}
\begin{aligned}
    a_k &= \begin{cases} 
        (-1)^k \frac{2}{\pi k} & k > 1 \\
        1 & k = 1
    \end{cases}
\end{aligned}
\end{equation}
to synthesize a (band-limited) sawtooth-like waveform with lower amplitudes towards higher partials.

Aliasing-free synthesis requires the highest partial frequency to be below half the Nyquist frequency $f_\mathrm{s} / 2$. 
Hence, the highest possible number of partials $K_\mathrm{max}$ depends on the highest fundamental frequency value in the PC  
${f_{0, \mathrm{max}} = \underset{n}{\max} \bigl( f_0(n) \bigr)}$ and is given by
\begin{equation}
    K_\mathrm{max} = \bigg\lfloor 
    \frac{f_\mathrm{s}}{2 \cdot f_{0, \mathrm{max}}}
    \bigg\rfloor
\end{equation}
For the synthesis of each PC, we randomly sample the number of partials as ${K  \sim \mathcal{U}(1, K_\mathrm{max})}$.

\subsection{\label{sec:spc_pitch_contour} Discussion}

The seven PC types, whose parameters are sampled within empirically determined value ranges accounting for different audio domains, represent somewhat idealized contour shapes. Although these generated PCs can approximate natural PCs on a local scale, certain limitations exist in their expressiveness.
First, in the PC model introduced in \secref{sec:contour_model}, the slope of the glissando PCs is constant, even though this parameter is likely tempo-dependent in musical contexts.
In addition, the \textit{bend} PCs have a temporally symmetrical shape, although, for example, the release part of the contour often takes longer for bendings played on the guitar.
Finally, as discussed in \secref{sec:audio_synthesis}, we limit our PC synthesis to a sawtooth-like waveform, which does not fully capture the physical complexities of time-varying amplitudes of individual partials or the inharmonic relationships between partial frequencies.

Finally, as discussed in \secref{sec:spc_pitch_contour}, we restrict ourselves to a sawtooth-like waveform for the synthesized PC signals, which does not reflect the physical reality of time-varying amplitudes of individual partials or inharmonic partial frequency relationships.


\section{\label{sec:e2e_pc_modeling}Pitch-Tracking-Based Contour Analysis}


\subsection{\label{sec:pitch_tracking} Pitch-Tracking Algorithms}

Traditionally, the first step to classify the type of a PC is to estimate the F0 trajectory from an audio signal using a pitch-tracking algorithm.
These algorithms generally rely on several assumptions, such as the  audio signal to be monophonic (single-pitch) at a low noise level and the PC to have a smooth F0 trajectory without abrupt changes.
Given the various contour shapes and parameter ranges mentioned in the study, designing a pitch-tracking algorithm that works effectively across various audio domains is a major challenge.
For example, the expected rate of change (modulation frequency) in the F0 trajectory sets an upper limit on the frame size, in which the F0 is assumed to be constant.
This, in turn, limits the frequency resolution due to the fundamental time--frequency uncertainty.
In our study, we evaluate the following well-established pitch-tracking algorithms, which are based on signal processing or deep learning. 



\texttt{pYIN} \cite{Mauch_2014_pYin_ICASSP} is a time-domain 
pitch-tracking algorithm based on pitch candidates extracted from the auto-correlation function. 
A post-processing step selects the most likely pitch candidate based on model assumptions like smoothness and pitch range, tailored for musical instruments and voice.

\texttt{SWIPE} \cite{CamachoH08_SawtoothWaveform_JASA} is a frequency-domain algorithm that relies on the correlation of the spectral frames with carefully designed kernels that represent individual pitch candidates. Different DFT window sizes are used for different pitch candidates to improve the matching quality between kernels and spectral frames. The approach avoids heuristics to smooth frame trajectories, making all model assumptions explicit in the kernel design

\texttt{CREPE} \cite{Wook_2018_CREPE_ICASSP} is an end-to-end model for pitch-tracking based on a CNN. 
The model has a limited F0 range up to \SI{1975.5}{Hz} since the
training data include only musical instrument and singing recordings.
The restricted frequency range poses a significant constraint for cross-domain pitch analysis.
Simply allowing for a wider range of possible pitches would not only considerably increase the model complexity, but would also require an even larger and more diverse training dataset.


\textbf{Evaluation on SPC dataset:} 
In order to select the most suitable pitch-tracking method for a cross-domain PC analysis, we first compared the pitch-tracking performance of all three methods on the SPC dataset.
We used the raw pitch accuracy metric RPA50 to measure the fraction of time frames, where the estimated F0 is within 50 cents of the ground truth value \cite{BittnerB19_F0Evaluation_ISMIR, RaffelMHSNLE14_MirEval_ISMIR}. 

\begin{figure}[t]
\includegraphics[width=0.5\textwidth]{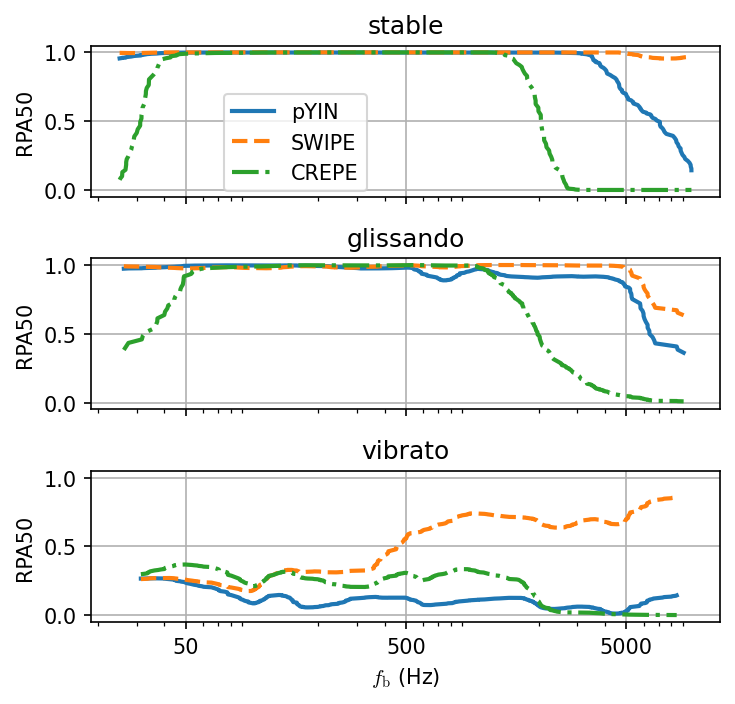}
\caption{\label{fig:results_pitch_tracking}
Results for the \texttt{pYIN}, \texttt{SWIPE}, and \texttt{CREPE} pitch-tracking algorithms over \textit{stable}, \textit{glissando}, and \textit{vibrato} PCs in the SPC dataset. Raw pitch accuracy (RPA50) is shown over the contour base frequency $f_\mathrm{b}$.}

\end{figure}

\figref{fig:results_pitch_tracking} illustrates the pitch-tracking performance of the three algorithms on \textit{stable}, \textit{glissando}, and \textit{vibrato} PCs as a function of the base frequency $f_\mathrm{b}$.
The graphs were obtained after a temporal smoothing using a 75-point Hann window moving average.
For \texttt{pYIN} and \texttt{SWIPE}, we use implementations provided by \cite{RosenzweigSM22_libf0_ISMIR-LBD}, and for \texttt{CREPE}, we use the pre-trained model provided by the original authors \cite{Wook_2018_CREPE_ICASSP}.

The results underscore each algorithm's limitations across various PC types and parameter ranges in the SPC dataset.
\texttt{pYIN} achieves the lowest RPA50 for fluctuating \textit{vibrato} PCs, due to the aforementioned model assumptions in its post-processing.
The performance of \texttt{CREPE} confirms its limitation to the frequency range of musical instruments.
Using \texttt{SWIPE} results in the highest overall RPA50 on the SPC dataset. In particular, it outperforms the other algorithms for \textit{vibrato} PCs towards higher base frequencies.
Note that \texttt{SWIPE} is specifically designed for sawtooth waveforms, so that it is expected to perform better for the SPC dataset compared to signals with other waveforms or timbres.


Due to its superior performance, we selected \texttt{SWIPE} as  pitch-tracking algorithm for the \texttt{PT-1D} approach.
Given an audio clip, \texttt{SWIPE} is used to calculate frame-wise F0 estimates, which serve as input to a DNN model with a convolutional front-end, which will be introduced in \secref{sec:nn_architecture_e2e}.
For \texttt{SWIPE}, we use a hopsize of 48 samples, yielding 1000 frame-level F0 estimates for a one-second long PC. 
As we have empirically found that higher strength threshold values impair the pitch-tracking performance, we use a threshold of zero.
Furthermore, we adapt \texttt{SWIPE} to cover the entire frequency range of the SPC dataset from \SI{25}{Hz} to \SI{10}{kHz} with a frequency resolution of \SI{128}{bins} per octave.



In the experiments described in \secref{sec:evaluation}, we additionally use the ground truth F0 contours of the SPC dataset as \texttt{ORACLE} input representation for the \texttt{PT-1D} approach.
Using this representation, we can quantify the influence of potential pitch estimation errors introduced by \texttt{SWIPE}.
To restrict the input features to a reasonable value range, both the \texttt{SWIPE} and \texttt{ORACLE} PCs are normalized by the mean and mean standard deviation that is calculated over the whole training dataset.

\subsection{\label{sec:nn_architecture_e2e} Neural Network Front-End}

For the end-to-end approach of directly estimating the PC parameters from \texttt{SWIPE} or \texttt{ORACLE}, we use a 1D convolutional neural network.
It consists of four layers with 512, 256, 128 and 64 filters, respectively, all with filter length 16 and a stride length of 4.
Given the depth of the network and width of convolutional filters, this results in a theoretical receptive field of 5101 frames in the input.
We found experimentally that having a larger receptive field than the 1000 required frames results in better performance of the model.
The given configuration results in a total of 3.75M trainable parameters, comparable to the architecture used for the vision-based approach below.



\section{\label{sec:vb_pc_modeling} Vision-Based Pitch Contour Analysis}

In computer vision (CV), a wide range of deep neural network models have been developed for object classification in natural images.
These models have in common that the complexity of the learned feature representations progressively increases with each layer (hierarchical feature learning) \cite{GoodfellowBC_2016_DeepLearning_MIT}.
Filters in early convolutional layers correspond to basic geometric shapes, such as curves and lines, whereas filters in later layers correspond to more complex shapes.
Our main assumption is that the PC types introduced in \secref{sec:contour_model} include curve and line elements similar to those found in natural images.
Therefore, 
we propose to process a time-frequency representation of an audio clip using a deep neural network that has been pre-trained for classifying natural images.

\subsection{\label{sec:tf_representations} Time--Frequency Representations}

\begin{figure}[t]
\includegraphics[width=.5\textwidth]{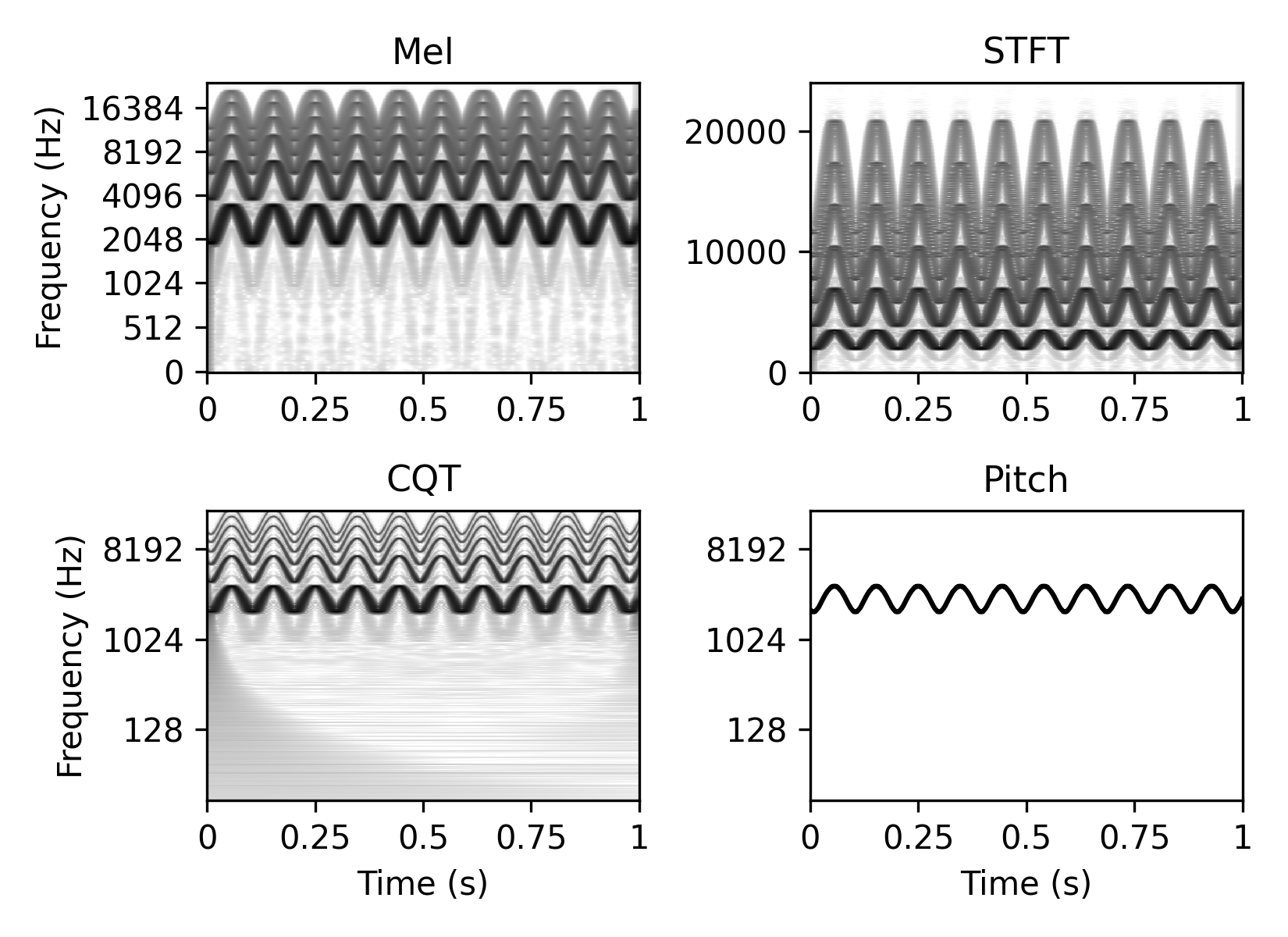}
\caption{\label{fig:example_feature}Four different feature representations extracted from the file \texttt{vibrato\_425.wav} of the SPC dataset with six partials and the contour parameters $f_\mathrm{b} = \SI{2696.6}{Hz}$, $f_\textrm{m}=\SI{10.3}{Hz}$, and ${\Delta_f=\SI{439.1}{cents}}$.}
\end{figure}

\figref{fig:example_feature} illustrates 
four different time--frequency representations 
of a \textit{vibrato} PC, which can be processed by the neural network.
In our study, we compare
a log-magnitude Mel spectrogram (denoted as \texttt{Mel}) with 128 Mel bands, a log-magnitude Constant-Q transform (\texttt{CQT}), and an Short-time Fourier Transform (\texttt{STFT}) log-magnitude spectrogram with a linearly spaced frequency axis.
As an additional fourth representation, we use a binary pitch--time representation (\texttt{Pitch})
that captures the ground-truth PC without any overtones. This oracle representation can considered as the output of an ideal, error-free pitch-tracking algorithm.
In accordance with the time resolution of the generated PCs in the SPC dataset, we consistently use a hop length of \SI{1}{ms} (\SI{48}{samples} at a sample rate of \SI{48}{kHz}) and an FFT size of 2048 samples.
The feature representations \texttt{CQT} and \texttt{Pitch} are based on a logarithmic frequency axis between \SI{25}{Hz} and \SI{20}{kHz} with a frequency resolution of 60 bins per octave (5 bins per semitone).



\subsection{\label{sec:nn_architecture_vb} Neural Network Front-End}

After a suitable time--frequency representation of the input audio is extracted, it is processed as image-like input feature by a CNN.
As network front-end, we use all layers except the final classification layer of a MobileNetV2 model \cite{Sandler_2018_MobileNetV2_CVPR}, which has been pre-trained for object detection in natural images using the ImageNet dataset \cite{Deng_2009_ImageNet_CVPR}.
We selected the MobileNetV2 model because it has a medium size of 3.5 million parameters, as well as the lowest (CPU-based) inference time among state-of-the-art computer vision models\footnote{\url{https://keras.io/api/applications/}} and achieved a competitive top-1 accuracy of $71.3\%$ on the ImageNet dataset.
The standard input format of the MobileNetV2 model is an image of size ${224 \times 224}$ pixels with three color channels.
Given an audio clip of arbitrary length, we first partition it into non-overlapping one-second-long segments, map each segment to an image of size 224 $\times$ 224 pixels using bicubic interpolation, replicate the feature values to three color channels, and finally normalize the entire batch of images extracted from a single audio clip to a range between -1 and 1.

\begin{table*}[t]
\caption{\label{tab:experiments_spc}
Accuracy ($A$) for PC type classification and mean absolute error (MAE) for regression of PC parameters $f_\mathrm{b}^\mathrm{cent}$, $\Delta_f$, and $f_\mathrm{m}$ using the SPC dataset. 
Results are shown for the end-to-end approach \texttt{PT-1D} (top) and the vision-based approach \texttt{VI-2D} (bottom) based on different input representation.
For the latter, metrics are given for \texttt{VI-2D-I} and for \texttt{VI-2D-R} (in brackets).
Oracle representations ($^\circ$) are tested for both approaches (see \secref{sec:pitch_tracking} and \secref{sec:tf_representations}).
The best (non-oracle) results are shown in bold font for each metric.}

\begin{center}
\begin{tabular}{llrrrrrrrr}
\toprule
\textbf{Approach} & \textbf{Input}  & 
\multicolumn{2}{l}{$A$} &
\multicolumn{2}{l}{MAE($f_\mathrm{b}^\mathrm{cent}$)} &
\multicolumn{2}{l}{MAE($\Delta_f$)} &
\multicolumn{2}{l}{MAE($f_\mathrm{m}$)} \\
\midrule
\texttt{PT-1D} & \texttt{SWIPE}  & 0.68 & & 169.0 & & 89.6 & & 4.93 \\
    & \texttt{ORACLE}$^\circ$ & 0.82 & & 82.8 & & 35.7 & & 1.09 \\
\midrule
\texttt{VI-2D} & \texttt{Mel}  & \textbf{0.93} & (0.58) & 133.8 & (200.0) & 38.3 & (93.8) & \textbf{2.51} & (6.50) \\
   & \texttt{CQT}  & \textbf{0.93} & (0.68) & \textbf{114.8} & (227.3) & \textbf{36.6} &(83.6) & 2.80 & (6.45) \\
   & \texttt{STFT}  & 0.89 & (0.60) & 165.8 & (276.3) & 47.7 & (115.1) & 3.24 & (7.10) \\
   & \texttt{Pitch}$^\circ$ & 0.86 & (0.67) & 77.6 & (1343.3) & 25.2 & (260.3) & 4.21 & (7.31) \\
\bottomrule

\end{tabular}
\end{center}
\end{table*}

\section{Model Pre-Training on SPC Dataset}
\label{sec:evaluation}

As illustrated in \figref{fig:overall_flowchart}, 
we follow a transfer learning strategy for both PC modeling approaches \texttt{PT-1D} and  \texttt{VI-2D} models introduced in \secref{sec:e2e_pc_modeling} and \secref{sec:vb_pc_modeling}.
As a first step, the models are pre-trained on the SPC dataset.
As a second step, the pre-trained models are fine-tuned on specific downstream classification tasks for cross-domain PC analysis (see \secref{sec:downstream_tasks}).
In this section, we will focus on the first step, where each model is trained on joint classification and regression tasks related to different PC parameters. 

\begin{figure}[t]
  \centering
  \subfigure[Model back-end \& loss functions.]{
    \includegraphics[width=.8\linewidth]{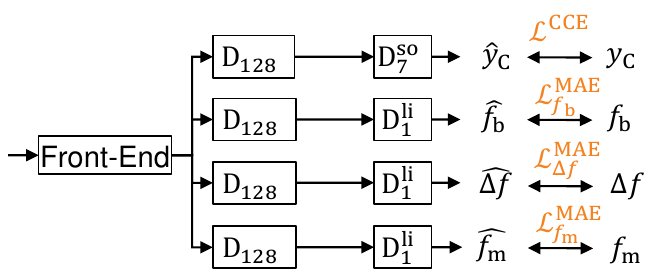}
    \label{fig:eval_prot_train}
    }
    \\ 
  \subfigure[Fine-tuning of pre-trained model back-end on downstream task $T_i$.]{  
    \includegraphics[width=.8\linewidth]{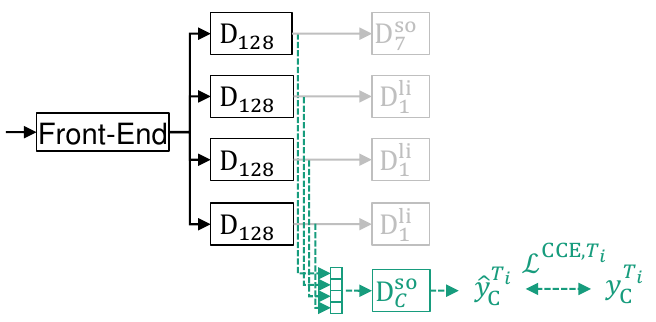}
    \label{fig:eval_prot_finetuning}
  }
  \caption{(a) Unified back-end architecture, which combines a model-specific front-end with four output heads for joint PC type classification (upper head) and PC parameter regression (second to fourth head). The predicted variables and the corresponding loss functions (orange) are provided for each model head and detailed in \secref{sec:eval_network}. (b) Embedding vector concatenation for fine-tuning on downstream tasks.}
  \label{fig:combined}
\end{figure}

\subsection{\label{sec:eval_network} Neural Network Back-End}

In both models, we use the same back-end architecture shown in \figref{fig:eval_prot_train}, which connects one of the two front-ends (see  \secref{sec:nn_architecture_e2e} and \secref{sec:nn_architecture_vb}) to four output heads, which implement the classification of the PC type (upper head) as well as the regression of the PC parameters base  
frequency $f_\mathrm{b}^\mathrm{cent}$, measured in cents relative to the lowest base frequency of \SI{25}{Hz} as
\begin{equation}
f_\mathrm{b}^\mathrm{cent} = 1200 \cdot \log_2 \left(f_\mathrm{b}/25\right),
\end{equation}
modulation extent $\Delta f$ in cents, and modulation frequency $f_\mathrm{m}$ in Hz (second to fourth head from the top). 

Each output head includes dense blocks (D), which are composed of a dense layer with the number of neurons denoted in the subscript, a dropout layer with ratio 0.2, and an activation function, which is either ReLU (no superscript), softmax (superscript ``so''), or linear (superscript ``li'').

\subsection{\label{sec:loss_functions} Loss Functions}

In the model training shown in \figref{fig:eval_prot_train}, we use
a multi-task loss function defined as a weighted combination of the categorical cross-entropy (CCE) loss function for PC type classification and the mean absolute error (MAE) loss function for each of the regression tasks, respectively: 
\begin{equation}
\mathcal{L}_{\mathrm{multi}} = 10 \mathcal{L}^\mathrm{CCE} + 0.1  \left(\mathcal{L}^\mathrm{MAE}_{f_\mathrm{b}^\mathrm{cent}} + \mathcal{L}^\mathrm{MAE}_{\Delta_f} + \mathcal{L}^\mathrm{MAE}_{f_\mathrm{m}} \right).
\end{equation}
\label{eq:loss}
After examining a variety of weighting factors, we discovered that the chosen loss weights optimally balanced the CCE and MAE loss components.
During training of the \texttt{VI-2D} model, we compare front-end weights pre-trained on ImageNet (\texttt{VI-2D-I}) with randomly initialized weights (\texttt{VI-2D-R}).


For the downstream tasks discussed in \secref{sec:downstream_tasks}, we require single-task classification models, which we obtain by modifying the pre-trained multi-task models as illustrated in \figref{fig:eval_prot_finetuning}.
In particular, we concatenate the embeddings within all four output heads prior to the final classification layers (green dashed lines) and add a single classification layer with a softmax function (green box) according to the number of classes in task $T_i$.
The single-task model is trained using the categorical crossentropy loss function as
\begin{equation}
\mathcal{L}_{\mathrm{single}} = \mathcal{L}^{\mathrm{CCE}, T_i}.
\end{equation}

Due to the individual training objectives combined in $\mathcal{L}_{\mathrm{multi}}$, we consider the concatenated embedding vector to provide better interpretability since its four segments represent the four PC properties overall shape, frequency position, modulation extent, and modulation rate. 


\subsection{\label{sec:embedding_extractor} Model Training}

We define a training/test split for the SPC dataset by using the first 400 examples of each PC type as training set and the remaining 100 examples as test set.
We train the multi-task models for $700$ epochs using the Adam optimizer \cite{Kingma_2015_adam_iclr}, a learning rate of $5 \cdot 10^{-4}$, and a batch size of $32$. 
As evaluation metrics, we compute the the multi-class accuracy $A$ to evaluate the PC type classification and the mean absolute error (MAE) to evaluate the three regression tasks.

\subsection{\label{sec:vision_based_contour_features} Results}

\tabref{tab:experiments_spc} summarizes the evaluation results on of the SPC data set.
When comparing the \texttt{SWIPE} and \texttt{ORACLE} representations for the \texttt{PT-1D} approach, we observe that pitch-tracking inaccuracies lead to diminished classification and regression performance across all PC parameters.
Notably, the results obtained from the \texttt{ORACLE} input of \texttt{PT-1D} demonstrate that the convolutional architecture performs better than the \texttt{VI-2D} approach for estimating local PC characteristics such as the modulation frequency $f_\mathrm{m}$ with an MAE of $1.09$ Hz.
At the same time, the performance is worse for global characteristics like the base frequency $f_\mathrm{b}^\mathrm{cent}$ (MAE values of $82.78$ vs. $77.6$).

The results for the \texttt{VI-2D} approach confirm that the proposed vision-based approach (\texttt{VI-2D}) effectively captures relevant PC characteristics related to shape, fundamental frequency, modulation extent, and modulation rate without any pitch-tracking step required.
In particular, the highest accuracy for PC type classification reaches $A=0.93$ for the \texttt{Mel} and \texttt{CQT} features.
A more detailed analysis showed that most confusions appear between the \textit{triangle} and \textit{sawtooth} classes (which have a similar PC shape) as well as the \textit{glissando} and \textit{stable} class (presumably for \textit{glissando} PCs with low modulation extent).

When comparing the \texttt{Mel} and \texttt{CQT} features on the regression tasks, we observe that the \texttt{CQT} feature provide a better performance with MAE values of around one semitone (114.8 cents) for the base frequency estimation ($f_\mathrm{b}^\mathrm{cent}$) and around a third of a semitone (36.6 cents) for the modulation frequency estimation ($\Delta_f$).
Finally, the modulation frequency ($f_\mathrm{m}$) is best estimated with an MAE of around 2.5 Hz using the \texttt{Mel} feature.

\subsection{Discussion}

Notably, the oracle representation (\texttt{Pitch}) performs poorly for PC type classification and regression of $f_\mathrm{m}$ in comparison but outperforms the other representations for the estimation of
$f_\mathrm{b}^\mathrm{cent}$ and $\Delta_f$.
We see two possible reasons for this. 
Given the limited resolution of 224 pixels along time and frequency, we assume that PCs with high modulation frequencies only appear blurred in the input representation. 
Compared to the other representations, the \texttt{Pitch} representation does not include additional frequency trajectories of the overtones (see \figref{fig:example_feature}).
These trajectories add redundancy to the feature representation, as they---while being at higher frequencies---resemble the F0 trajectory to be classified.
We assume that this redundancy facilitates the classification of contour types and the estimation of the modulation frequency $f_\mathrm{m}$.
This hypothesis is supported by the poor classification performance of the \texttt{PT-1D} approaches, where a small $\Delta_f$ can lead to negligible numerical changes in the normalized input, while the PC shape (e.g. of a bend) is exaggerated in the overtone contours.
In contrast, the additional overtones complicate the estimation of the base frequency $f_\mathrm{b}^\mathrm{cent}$ and the modulation range $\Delta_f$ due to ambiguities between harmonic components and the wider overall frequency bandwidth of the signal.

As general observation, we find that using front-end weights pre-trained on the ImageNet dataset (\texttt{VI-2D-I})  consistently leads to better results across all tasks and feature representations compared to randomly initialized weights (\texttt{VI-2D-R}).
Given these results, we use the \texttt{VI-2D-I} model and the \texttt{CQT} representation throughout the downstream task experiments that will be described in \secref{sec:downstream_tasks}.

\section{\label{sec:downstream_tasks} Model-Finetuning for Cross-Domain Downstream Tasks}

As a second stage of the overall transfer learning strategy, this section will discuss the fine-tuning of the pre-trained models (see \secref{sec:evaluation}) for several downstream classification tasks for which PCs are relevant.
We selected the corresponding datasets to cover various audio domains from music over speech and bioacoustics to everyday sounds and test the generalizability of the proposed vision-based approach \texttt{VI-2D} for cross-domain PC analysis.

\subsection{\label{sec:datasets} Datasets \& Data Splits}

Following the overview in \tabref{tab:datasets}, we first introduce all the downstream task datasets in this section.
If available, we used existing datasets with pre-defined training and test splits and created new splits if necessary. 
We will document all applied dataset splits on an accompanying website.
To provide a better impression of the audio characteristics in each dataset, we illustrate example CQT patches in \figref{fig:downstream_task_examples} for each dataset \texttt{VN}, \texttt{IB}, \texttt{EX}, and \texttt{EM} (top row, from left to right), as well as \texttt{EN}, \texttt{WA}, \texttt{ES}, and \texttt{RS} (bottom row, from left to right).

\textbf{Music: }
The \textbf{VocalNotes} \cite{Proutskova_2023_VocalNotes_ISMIR} dataset (\texttt{VN}) includes 10 minutes of vocal recordings from each of the five vocal traditions Japanese Min’yo, Chinese Bangzi opera,
traditional Russian village singing, Alpine yodelling, and Jewish Romaniote chanting.
Our focus is to classify the vocal tradition from a given PC.
The original authors confirmed that no official train--test split exists for this task.
Hence, we define a split in which a maximum of 50~\% of the recordings per class are used for the training set and the remaining recordings for the test set.
In this split (to avoid the album effect \cite{Mandel_2005_MusicClassification_ISMIR}, we assure that different parts and versions of the same songs are assigned to the same set.
The \textbf{IDMT-SMT-Bass} dataset \cite{Abesser_2010_BassStyles_ICASSP} (\texttt{IB}) 
is a benchmark data set to estimate the expression and plucking styles in the performance of electric bass guitar. The dataset includes isolated note recordings.
In this experiment, we focus on a set of 936 recordings of the six expression styles 
quarter-tone bending (BEQ), semi-tone bending (SEQ), slide down (SLD), slide up (SLU), fast vibrato (VIF), and slow vibrato (VIS), which were used in \cite{Abesser_2011_BassModulation_AES} as an extension of this dataset.
These recordings come from the same bass guitar but with three different pick-up settings. 
We use the recordings from the first setting (``BS\_1\_EQ\_1...'') as training data and the remaining recordings as test data.
This six-class taxonomy is closely related to the PC types and parameters introduced in \secref{sec:spc}. 
The low fundamental frequency range between around 40~Hz and 200~Hz poses an additional challenge to PC analysis using the \texttt{CQT} feature.

\begin{table*}[t!]
\caption{\label{tab:datasets}Selected downstream tasks related to PC analysis covering datasets from the four audio domains music, speech, bioacoustics, and everyday sounds. Last four columns provide the number of classes and macro-weighted F-score results for \texttt{VI-2D} model variants pre-trained on SPC dataset (\texttt{VI-2D-S}) or ImageNet dataset (\texttt{VI-2D-I}) and the \texttt{PT-1D} model.}


\begin{tabular}{p{.1\textwidth}p{.23\textwidth}p{.15\textwidth}p{.07\textwidth}p{.09\textwidth}p{.09\textwidth}p{.09\textwidth}}
\toprule
\textbf{Domain} & \textbf{Dataset} & \textbf{Task} & \textbf{Classes} & \multicolumn{3}{c}{$F$}\\
& & & & \texttt{VI-2D-S} & \texttt{VI-2D-I} & \texttt{PT-1D}\\
\midrule
Music & VocalNotes (\texttt{VN})\footnote{\url{https://zenodo.org/records/10065955}}
& Singing Style & 5  & 0.66 & 0.57 & 0.67  \\
      & IDMT-SMT-Bass (\texttt{IB})\footnote{\url{https://zenodo.org/records/7188892}} & Expression Style & 5  & 0.55 & 0.54 & 0.30 \\
\midrule
Speech & Expresso (\texttt{EX})\footnote{\url{https://speechbot.github.io/expresso/}}  & Style & 26 & 0.56 & 0.56 & 0.36 \\
       & EmoV\_DB (\texttt{EM})\footnote{\url{https://openslr.org/115/}}  & Emotion & 5 & 0.58 & 0.47  & 0.38 \\
\midrule
Bioacoustics & Enabirds (\texttt{EN})\footnote{\url{https://github.com/earthspecies/beans}} & Bird species & 34 & 0.29 & 0.27 & 0.09 \\
             & Watkins (\texttt{WA})\footnote{\url{https://github.com/earthspecies/beans}}\footnote{\url{https://whoicf2.whoi.edu/science/B/whalesounds/index.cfm}} & Marine mammals & 31 & 0.90 & 0.86 & 0.38 \\
\midrule
Everyday  & ESC-50 (\texttt{ES})\footnote{\url{https://github.com/karolpiczak/ESC-50}} & Sound event tagging & 50 & 0.68 & 0.73 & 0.16 \\
                Sounds & Regional Siren Dataset (\texttt{RS}) & Country of origin  & 9 & 0.72  & 0.63 & 0.38 \\
\bottomrule
\end{tabular}
\end{table*}
\textbf{Speech:}
As the first downstream task in the speech domain, we evaluate the classification of seven styles of expressive reading using audio recordings from the \textbf{Expresso} dataset \cite{Nguyen_2023_Expresso_INTERSPEECH} (\texttt{EX}).
For simplicity, we restrict our experiment to the short sentences (``base'' subset) extracted from the voice recordings read out (``read'' set). We  use recordings from the speakers ``ex01'' and ``ex02'' as training set and the remaining two speakers as test set. We consider the following eight expressive styles: confused, default, enunciated, happy, laughing, narration, sad, and whisper.
The \textbf{Emotional Voices Database}  \cite{Adigwe_2018_EmoVoicesDatabase_ARXIV}  (\texttt{EM}) includes 6893 English and French voice recordings spoken by two male and two female speakers which cover the following five emotions: amusement, anger, disgust, sleepiness, and neutral.
We focus on emotion classification and split the recordings on a speaker level such that both the training set and the test set include a male and a female speaker, respectively.

\textbf{Bioacoustics:}
In order to capture the wide variety of animal vocalizations with characteristic PCs, we select the \textbf{ena\-birds}  dataset \cite{Chronister_2021_Enabirds_ECO} (\texttt{EN}), which includes audio recordings with annotated segments of 33 bird species, and the \textbf{watkins}  dataset \cite{Sayigh_2016_WatkinsDB_MOA} (\texttt{WA}), which includes audio recordings of 31 subspecies of marine mammals such as dolphins, whales, and seals from the Benchmark of Animal Sounds (BEANS) \cite{Hagiwara_2022_BEANS_ARXIV}. Both datasets include animal vocalizations with characteristic PCs and provide pre-defined training and test splits.


\textbf{Everyday Sounds:}
The \textbf{ESC50} dataset \cite{Piczak_2015_ESC50_ACM} (\texttt{ES}) is a popular small benchmark dataset for environmental sound classification. It comprises \num{2000} audio clips (each five seconds of duration) from 50 sound classes that cover human and animal vocalizations, urban and domestic sounds, and natural soundscapes.
Although not all sound classes have characteristic PCs but rather noisy or transient spectra, we did not restrict the sound classes, but instead used the full dataset. 
A five-fold cross-validation split is provided for the dataset, of which we use the first fold as test set and the other folds as training set.
As a second dataset in this category, we use the \textbf{Regional Siren Classification}  dataset\footnote{\url{https://github.com/jakobabesser/regional_siren_classification_dataset}} (\texttt{RS}) which comprises emergency vehicle siren sounds from various countries. The dataset includes 270 labeled segments with 10 examples for each of 9 countries and 3 siren types (ambulance, firefighters, and police).
We create a simple 50-50 split into training and test subsets and aim at classifying the country of origin.

\subsection{Classification based on Confidence Weighting}
\label{sec:classification_procedure}

All downstream tasks discussed before are single-label multi-class classification tasks that are performed on audio clips, which are either complete audio files or segments thereof. 
In our experiments, we divide each audio clip into one-second long patches which are individually processed by the models.

In order to aggregate the patch-level model outputs to a clip-level result, we apply a confidence weighting scheme to mitigate the influence of silent or unvoiced patches without any PC.
Given a downstream task with $C$ classes and an audio clip split into $P$ patches, we denote the predicted class probabilities for the $k$-th patch as $p^{(k)}\in \mathbb{R}^{C}$.
We denote the class index for the highest value in $p_k$ as 
\begin{equation}
    c_1 = \arg\max(p^{(k)})
\end{equation}
and index of the second highest value as 
\begin{equation}
    c_2 = \arg\max_{\substack{j \in \left[1: C \right] \\ j \neq c_k}} p^{(k)}_{j}
\end{equation}
and measure the classification confidence $\alpha_k \leq 1$ as
\begin{equation}
    \alpha_k = p^{(k)}_{c_1}-p^{(k)}_{c_2}.
\end{equation}
Using a confidence weighting, we compute a pseudo-probability $\beta\in \mathbb{R}^{C}$ for each class as
\begin{equation}
    \beta_c = \frac{1}{P} \sum_{k=1}^{P} \alpha_k \cdot p^{(k)}_{c}
\end{equation}
for all classes $c \in \left[1:C \right]$ to finally obtain the predicted class index $\hat{y}$ as 
\begin{equation}
\hat{y} = \arg\max_{c \in \left[1: C \right]}\beta_c.
\end{equation}
The underlying assumption of the confidence weighting is that noise-like patches for some patch index $k \in \left[1:P\right]$ lead to an almost flat probability distribution $p^{(k)}$ over all classes and hence a low confidence value $\alpha_k$.

\subsection{\label{sec:model_configurations} Model Configurations}

In this section, we evaluate three approaches for PC analysis for eight downstream tasks.
On the one hand, we test the \texttt{VI-2D} approach either pre-trained on the ImageNet (\texttt{VI-2D-I}) or pre-trained successively on ImageNet and then on SPC (\texttt{VI-2D-S}) as explained in \secref{sec:evaluation}.
On the other hand, we evaluate the \texttt{PT-1D} with its convolutional front-end being pre-trained on the \texttt{ORACLE} PCs of the SPC dataset. All PCs are normalized to zero mean and unit variance using mean and standard deviation statistics computed over the entire SPC training dataset.

For each downstream task, we start from the initial weights of the pre-trained models and fine-tune them using on the training set of the corresponding dataset. In the fine-tuning process, all network parameters, including both front-end and back-end, are optimized. 
During the fine-tuning, we use a learning rate of $10^{-5}$ and 250 epochs for the \texttt{VI-2D} models and 500 epochs for the \texttt{PT-1D} model.


\subsection{\label{sec:results} Results}

\begin{figure}[t]
\includegraphics[width=.48\textwidth]{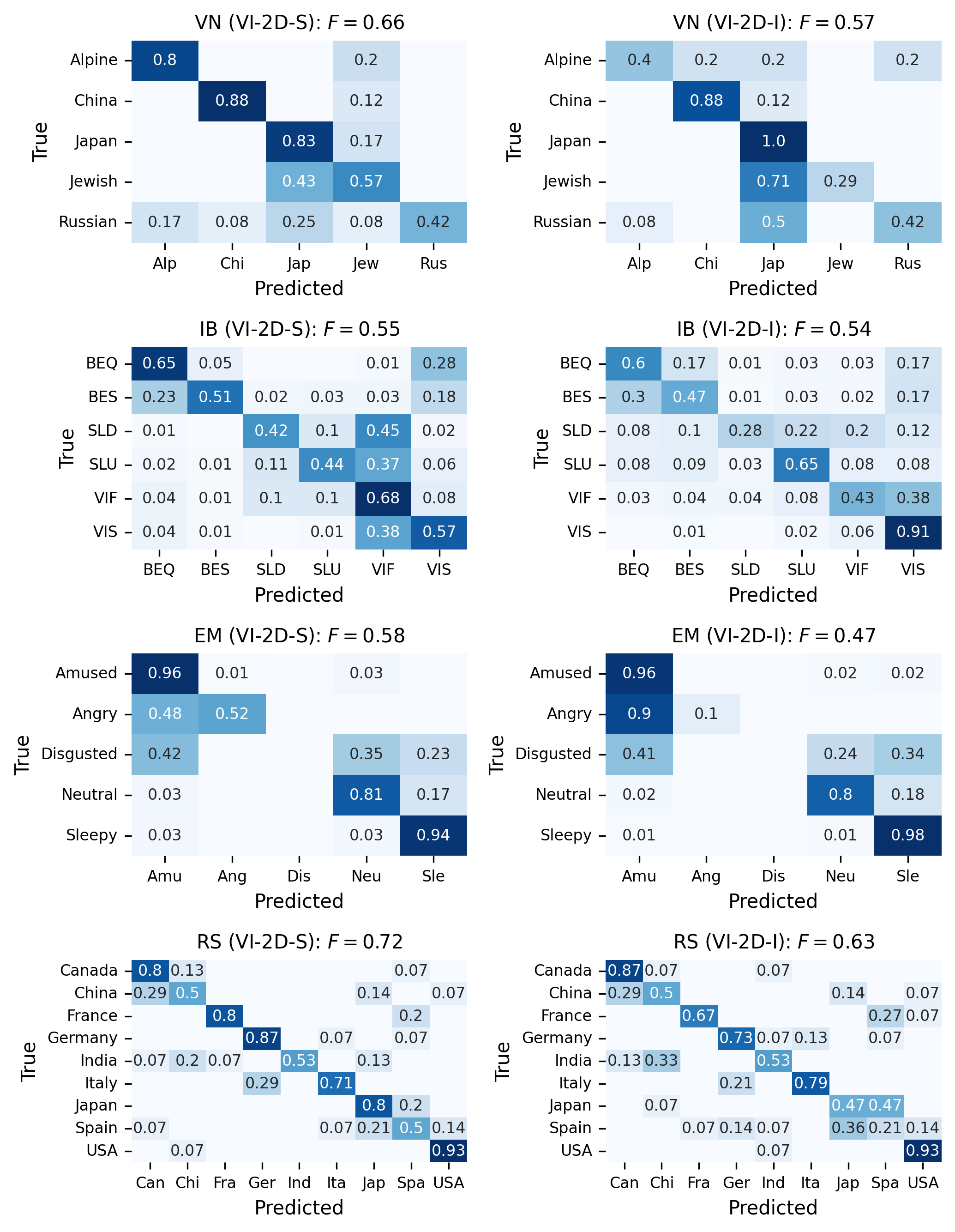}
\caption{\label{fig:downstream_confmat} Confusion matrices and F1 scores for selected downstream tasks contrasting the \texttt{VI-2D-S} approach (left column) and the \texttt{VI-2D-I} approach (right column).}
\end{figure}

In addition to the summary of each dataset in terms of downstream tasks and class numbers, \tabref{tab:datasets} shows macro-weighted F1 scores for three PC analysis approaches as the evaluation metric.
Given a binary classification problem, let TP, FP, and TN denote the number of true positives, false positives, and true negatives, respectively. 
The F1 score is computed as $F = {(2 \cdot TP)}/{(2 \cdot TP + FP + TN)}$.
All downstream tasks are multiclass classifications. The macro-weighted F1 score is the average of class-wise F1 scores for binary one-versus-all dataset partitions and adjusts for class imbalance.
In addition, in \figref{fig:downstream_confmat} we illustrate the confusion matrices obtained using the \texttt{VI-2D-S} and \texttt{VI-2D-I} models for a subset of the four datasets \texttt{VN}, \texttt{IB}, \texttt{EM}, and \texttt{RS} for better insight into the types of classification errors of both approaches.

For the music datasets \texttt{VN} and \texttt{IB}, we observe a higher F1 score for the \texttt{VI-2D-S} model than for the \texttt{VI-2D-I} model, which demonstrates that pre-training on synthetic PCs improves the classification results. 
In particular for the \texttt{VN} dataset, the F1 score of 0.66 for the \texttt{SPC} model confirms that the
different singing traditions exhibit some unique PC characteristics.
However, Russian village singing seems to share similar PCs with the other traditions as can be seen in the confusion matrices in \figref{fig:downstream_confmat} (first row).
For the \texttt{IB} dataset (second row), the improvement of the \texttt{VI-2D-S} approach compared to the \texttt{VI-2D-I} approach is only marginal, which we attribute to the low pitch range, in which the CQT provides only limited frequency resolution to distinguish between the subtle differences of the considered PC types.

In the speech domain, the \texttt{VI-2D-S} approach is only outperforming \texttt{VI-2D-I} for emotion recognition (\texttt{EM} dataset) but not for vocal style classification (\texttt{EX} dataset). This confirms that the emotional state of a speaker is directly linked to the pitch characteristics in speech \cite{Rodero_2011_IntonationEmotionSpeech_JV}.
As can be observed in the third row \figref{fig:downstream_confmat}, all emotions except ``disgusted'' seem to have somewhat unique PC shapes. As an exception, we observe a strong confusion between the ``amused'' and ``angry'' speech recordings, which could indicate that both have similar PCs.

In the bioacoustic domain, 
both datasets have a comparable class diversity with 34 bird species in the \texttt{EN} dataset and 31 marine mammal species in the \texttt{WA} dataset.
A notable difference is that while the \texttt{WA} dataset includes mainly the vocalizations of individual species, the recordings in the \texttt{EN} dataset capture the dawn chorus with significant temporal overlap between multiple bird species.
These different levels of difficulty are reflected in the F1 scores of up to $0.9$ for \texttt{WA} and only up to $0.29$ for \texttt{EN}.
In particular, the accuracy on the \texttt{EN} dataset is $0.62$, which confirms that the dataset has highly imbalanced classes. 
In general, the findings verify that specific marine mammal species exhibit highly distinctive PCs.

Finally, in the domain of everyday sounds, the importance of PC shapes in the regional classification of sirens is evident in the F1 score of $0.72$ of the \texttt{VI-2D-S} model for the \texttt{RS} dataset.
As shown in \figref{fig:downstream_confmat} (last row), the countries with the most unique PCs are Canada, France, Germany, Japan, and the USA.
Notably, when considering a larger set of 50 sound classes in the \texttt{ES} dataset, the more generic \texttt{VI-2D-I} model clearly outperforms the \texttt{VI-2D-S} model.
We hypothesize that PCs are only relevant for a subset of the sound classes, many of which, in contrast, are characterized more by transient or noisy components.
Pre-training only on the ImageNet dataset, which includes natural images with various shapes and textures, seems to be beneficial for recognizing more generic sets of sound classes.

The \texttt{PT-1D} model showed comparable performance to the \texttt{VI-2D} models only for the singing voice recordings in the \texttt{VN} dataset.
Apparently, the SWIPE pitch-tracking method is less reliable for lower  (\texttt{IB} dataset) and higher fundamental frequency ranges (e.\,g., \texttt{EN} and \texttt{WA} datasets).
Additionally, the pitch-tracking task is naturally ill-defined for non-voiced signal frames for example in speech consonants (\texttt{EX} and \texttt{EN} datasets) or in non-harmonic sound events (\texttt{ES}), for which no reasonable classification result can be expected.

\section{\label{sec:conclusion} Conclusions}

In this study, we aimed to investigate pitch contours as a unifying semantic concept, shared across diverse audio domains, such as music, speech, bioacoustics, and everyday sounds. 
Existing pitch-tracking methods are mostly optimized for music and speech, but face challenges particularly with wider frequency ranges and rapid pitch changes. 
As the main contribution of this work, we introduced an alternative methodology for pitch contour analysis that circumvents the necessity for a pitch-tracking step. We instead propose to process basic spectrogram-like representations of the audio signal by a deep convolutional neural network. 
In the first part of our study, we introduced the Synthetic Pitch Contour (SPC) dataset, comprising a large variety of synthetic pitch contours.
During an initial evaluation, the SWIPE algorithm outperformed other pitch-tracking methods, in particular, excelling at higher frequencies and rapid modulations.
The vision-based approach outperformed traditional pitch-tracking-based methods for pitch contour type classification and parameter regression.
Furthermore, model pre-training on object recognition in natural images further improved the model performance.

In the second part of our study, we evaluated both approaches on a diverse set of downstream classification tasks spanning four audio domains. 
The results show that state-of-the-art pitch-tracking algorithms do not deliver robust pitch contours across domains. 
In contrast, the vision-based PC analysis approach demonstrated superior performance, particularly after being further fine-tuned on the SPC dataset. 
In summary, the two-stage transfer learning approach consisting of pre-training on the ImageNet dataset and the SPC dataset and subsequent fine-tuning allows the model to specialize on downstream classification tasks.

By introducing a novel method for pitch contour analysis without explicit pitch-tracking required, this study lays the foundation for further comparative studies of pitch contour properties across various audio domains.
In future work, we plan to analyze the temporal structure of longer pitch contours and study pitch contour parameters in specific audio domains, for example, to compare vocalizations across different animal species.

\section*{\label{sec:acknowledgements} Acknowledgements}

The authors thank Judith Bauer for fruitful discussions.
This work was funded by the Deutsche Forschungsgemeinschaft (DFG, German Research Foundation) under Grant Nos. 350953655 (MU 2686/11-2, AB 675/2-2) and 401198673 (MU 2686/13-2).
The authors are with the International Audio Laboratories Erlangen, a joint institution of the Friedrich Alexander-Universität Erlangen-Nürnberg (FAU) and Fraunhofer Institute for Integrated Circuits IIS. The first author's work was furthermore funded by the EU Horizon Europe research and innovation program (Grant No. 101081964) and the vera.ai project (Grant No. 101070093).

\bibliographystyle{IEEEtran}  
\bibliography{idmt_references, idmt_references_new, references_new, references_jakob}



\end{document}